\documentclass[aps,twocolumn,floats,superscriptaddress,prd,nofootinbib]{revtex4-1}
\pdfoutput=1

\usepackage{amsmath}
\usepackage{amssymb}
\usepackage{graphicx}
\usepackage{epstopdf}
\usepackage{hyperref}

\usepackage{feynmp}
\makeatletter
\def\endfmffile{%
  \fmfcmd{\p@rcent\space the end.^^J%
          end.^^J%
          endinput;}%
  \if@fmfio
    \immediate\closeout\@outfmf
  \fi
  \ifnum\pdfshellescape>\z@
    \immediate\write18{mpost \thefmffile}%
  \fi}
\makeatother
\newcommand{\dif}{\mathrm{d}}

\begin{document}

\title{Perturbative Unitarity of Inflationary Models with Features}

\author{Dario Cannone}
\email{dario.cannone@pd.infn.it}
\author{Nicola Bartolo}
\email{nicola.bartolo@pd.infn.it}
\author{Sabino Matarrese}
\email{sabino.matarrese@pd.infn.it}
\affiliation{Dipartimento di Fisica e Astronomia ``G. Galilei'', Universit\`a degli studi di Padova, via Marzolo 8, I-35131, Padova, Italy}
\affiliation{INFN, Sezione di Padova, via Marzolo 8, I-35131, Padova, Italy}

\newcommand{\imgref}[1]{figure \ref{#1}}

\begin{abstract}
We consider the pertubative consistency of inflationary models with features with effective field theory methods.
By estimating the size of one-loop contributions to the three-point function, we find the energy scale where their contribution is of the
same order of the tree-level amplitude. It is well-known that beyond that scale, perturbative unitarity is lost and the theory is no more
under theoretical control. Requiring that all the relevant energy scales of the problem are below this cutoff, we derive a strong upper
bound on the sharpness of the feature, or equivalently on its characteristic time scale, which is independent on the amplitude of the
feature itself. We point out that the sharp features which seem to provide better fits to the CMB power spectrum are already
outside this bound, questioning the consistency of the models that predict them.
\end{abstract}

\maketitle

\section{Introduction}

The recent analysis of \emph{Planck} data does not show any significant deviations from the simplest single-field slow-roll models of
inflation, but some issues remain open. In particular, it seems that there could be a relative better fit to the curvature power spectrum
if the possibility of small and rapid oscillations is taken into account \cite{Planck_ps, Hu2013, Benetti2013, Achucarro2013}. Even though,
from present data, it seems that there is not enough improvement in the fits to assess their statistical significance
\cite{Spergel2013a, Spergel2013b}, it is very interesting to study the models that could provide a primordial origin to these
signatures.
In this paper we will focus on the so-called ``models with features'', which are well-known in the literature \cite{Starobinsky1992, Wang1999,
Adams2001, Peiris2003, Covi2006, Chen2006b, Covi2007, Chen2008, Piazza2008, Dvorkin2009, Hazra2010, Adshead2011, Adshead2011b, Chen2011a,
Arroja2011b, Arroja2012, Miranda2012}. Tipically, oscillations in the power spectrum are due to some features in the potential of the
inflaton or in the speed of sound, which induce a
temporary deviation from the slow-roll dynamics that comes back to the attractor solution in less than few efolds
without ruining inflation.
Very interestingly, features also induce non-Gaussianity, giving us the possiblity to constraint
these models also with an other observable, the bispectrum \cite{planck_bs}.

In this context, from a theoretical point of view, a useful tool can be the Effective Field Theory of Inflation
(EFTI) \cite{Creminelli2007, Weinberg2008}. Independently of the mechanism that could produce the features, their effects
on observables can equivalently be described through the time-dependent coefficients of the effective action of the Goldstone boson
that non-linearly realizes time-diffeomorphisms,
\begin{eqnarray}\label{S_eft}
 S = & \displaystyle \int \dif^4 x \sqrt{-g} & \left[-M_{Pl}^2\dot{H}(t+\pi)\left(\dot{\pi}^2-\frac{{(\partial_i \pi)}^2}{a^2}\right) \right.+\nonumber \\
   & & +\,2M_2^4(t+\pi)\left(\dot{\pi}^2 + \dot{\pi}^3 -\dot{\pi} \frac{{(\partial_i \pi)}^2}{a^2}\right)- \nonumber \\
   & & \left. \frac{4}{3}M_3^4(t+\pi)\dot{\pi}^3+  .\,.\,. \right]\; ,
\end{eqnarray}
where $\pi$ is related to the curvature perturbation $\zeta$ by $\zeta=-H\pi$. Consider, for example, a time dependence of the form
\begin{equation}\label{H_feat}
  \dot{H}(t)=\dot{H}_0(t)\left[1+\epsilon_{step}\,F\left(\frac{t-t_f}{b}\right)\right] \; ,
\end{equation}
where the function $F$ represent a step centered in $t_f$ with a height $\epsilon_{step}$
and a characteristic width $\Delta t = b$. It has been shown that, when inserted into the effective action \eqref{S_eft},
one obtains the predicted damped oscillations in the power spectrum for models with features in the potential, simply neglecting the
$M_n$ coefficients \cite{Cannone2013}. Moreover the EFTI approach allow us to go beyond the standard scenario and generalize features in a model
independent way to the speed of sound, $c_s$, and any other coefficients.
Besides the spectrum, from the Taylor expansion
\begin{equation}
 \dot{H}(t+\pi)=\dot{H}(t)+\ddot{H}(t)\pi+ ... \; ,
\end{equation}
we get new interactions as, for example, the cubic term:
\begin{equation}
 \mathcal{L}_3\ni\,-M_{Pl}^2\ddot{H}(t)\,\pi\dot{\pi}^2 \; ,
\end{equation}
which gives the largest contribution to the three-point function in the case of sharp feature.
The resulting bispectrum is not scale-invariant and will be
peaked for those modes that at the time of the feature, $t_f$, have an energy comparable to the inverse of the characteristic time scale,
$b$, of the feature \cite{Cannone2013}. If we define a sharpness parameter
\begin{equation}
 \beta=\frac{1}{bH} \; ,
\end{equation}
as the ratio between the energy $1/b$ and the Hubble scale $H$,
it can be shown that the modes that are most affected are more inside horizon as the feature becomes sharper and sharper
\cite{Chen2008, Chen2011a, Adshead2011}.
Moreover, as the amplitude
at the peak grows quadratically with the sharpness, this can become the major source of non-Gaussianity and be possibly
seen in future data analyses. However, being $\beta$ unconstrained, non-linearities could also be too large, not only for
observations, but also for the theretical consistency of the models.

\section{Energy Scales and Unitarity}

The validity of the perturbative treatment one commonly uses relies on the assumption that higher-order contributions are smaller. This is
what is done for example when one computes the equations of motion truncating the action at second order: it is implicitly assumed
that the third-order contribution $\mathcal{L}_3$, for example, is small compared to the quadratic Lagrangian $\mathcal{L}_2$.
To confirm that assumption, then one should check
that $\mathcal{L}_3/\mathcal{L}_2\ll1$ in the relevant energy scales of the problem, so that the theory is perturbatively safe.
In the standard cases, the only relevant energy scale is $H$, where fluctuations are crossing the horizon, so the bound is taken at $E\sim H$.
However, for inflationary models with features (or resonances), this should be required also for the scale where the largest interaction happens
\cite{Behbahani2011, Cannone2013}, which corresponds to the inverse of the relevant time-scale $b$
of the feature (or the resonance).
In the case of inflationary models with features, we should make sure that $\mathcal{L}_3/\mathcal{L}_2\ll1$ is valid even in the worst
possible case i.e. at the time of the feature $t_f$, when the interaction is maximized. Given that, one can find \cite{Cannone2013}
\begin{equation}
 \frac{\mathcal{L}_3}{\mathcal{L}_2}\bigg|_{E\sim\beta H}\ll1\qquad \Longrightarrow \qquad
 \beta^2\lesssim\frac{1}{\epsilon_{step}\mathcal{P}_{\zeta,0}^{1/2}} \; ,
\end{equation}
where
\begin{equation} \label{P0}
 P_{\zeta,0}(k)=\frac{2\pi^2}{k^3}\mathcal{P}_{\zeta,0}=\frac{H^2}{4M_{Pl}^2\epsilon }\frac{1}{k^3}
\end{equation}

However we should check also that higher-order contributions from $\mathcal{L}_n$ satisfy a similar bound.
In order to do this, notice that the most important interaction in the Lagrangian at $n$th-order
(which comes from the Taylor expansion of the
term $\dot{H}(t+\pi)$ in the effective action \cite{Behbahani2011}), parametrically scales as
\begin{equation}
 \mathcal{L}_n\sim M_{Pl}^2H^{(n-1)}\pi^{n-2}\dot{\pi}^2 \; ,
\end{equation}
while
\begin{eqnarray}
 \dot{H} & \sim & \epsilon H^2 \; , \label{scaling1}\\
 H^{(n)} & \sim & \epsilon\, \epsilon_{step} \beta^{n-1} H^{n+1} \; . \label{scaling2}
\end{eqnarray}
Our perturbative expansion is then safe if:
\begin{equation} \label{Ln/L2}
 \frac{\mathcal{L}_n}{\mathcal{L}_2}\bigg|_{E\sim\beta H}\sim\epsilon_{step}\beta^{2n-4}\zeta^{n-2} \ll1 \; ,
\end{equation}
which implies
\begin{equation}
 \beta^2\lesssim\frac{\mathcal{P}_{\zeta,0}^{-1/2}}{\epsilon_{step}^{1/(n-2)}} \; \stackrel{n\gg1}{\sim} \; \mathcal{P}_{\zeta,0}^{-1/2} \; ,
\end{equation}
where in the last step we take the limit for $n\to\infty$.
This simple argument then suggests that we should take $\beta^2\lesssim\mathcal{P}_{\zeta,0}^{-1/2}$ if we do not want higher-order
corrections to threat perturbativity. An important thing to note here is that, being inside the horizon, our theory is a quantum theory,
so the violation of \eqref{Ln/L2}
is signaling an actual quantum-mechanical strong coupling (in the sense that quantum loops are not suppressed),
so that unitarity is lost and the model is not under control \cite{Creminelli2007, Shandera2008, Baumann2011, Baumann2011c}.
In order to state the problem more rigorously, we will estimate the amplitudes of one-loop contributions to the three-point function
and compare them to the tree-level amplitudes\footnote{Notice that one can obtain the same result considering, for example, one-loop
contributions the two-point instead of the three-point function.}. 

Consider the cubic operator,
\begin{equation}
\mathcal{L}_3\ni M_{Pl}^2 \, \ddot{H}\left(\frac{t-t_f}{b}\right)\,\pi \dot{\pi}^2 \; ,
\end{equation}
at the time of the feature, $t_f$, where the interaction is maximal.
Upon canonical normalization, $(-2M_{Pl}^2\dot{H})^{-1/2}\pi=\pi_c$, and using \eqref{scaling2}, we have:
\begin{equation}
 \frac{1}{2}\frac{\epsilon_{step}\beta}{M_{Pl}\sqrt{2\epsilon }}\pi_c\dot{\pi}_c^2=\epsilon_{step}\, g \,\pi_c\dot{\pi}_c^2 \; .
\end{equation}
Notice that, as the operator $\pi \dot{\pi}^2$ has mass-energy dimension $E^5$, the coupling $g$ in front of it has dimension $1/E$.
Diagrammatically, the corresponging vertex and amplitude (by dimensional analysis) are:

 
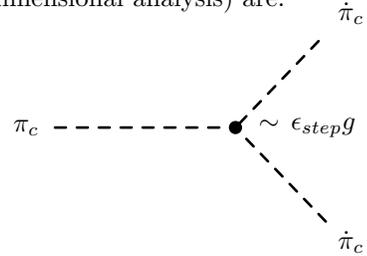
\begin{center}
 \begin{fmffile}{3p}
 \begin{fmfchar*}(40,25)
 \fmfleft{pi} \fmflabel{$\pi_c$}{pi}
 \fmf{dashes}{pi,3p}
 \fmflabel{$\; \sim\, \epsilon_{step} g$}{3p}
 \fmf{dashes}{dotpi1,3p,dotpi2}
 \fmfright{dotpi1,dotpi2} \fmflabel{$\dot{\pi}_c$}{dotpi1} \fmflabel{$\dot{\pi}_c$}{dotpi2}
 \fmfdot{3p}
 \end{fmfchar*}
 \end{fmffile}
 \end{center}
\begin{center}
 \begin{equation} \mathcal{M}^{(0)} \sim \epsilon_{step} \,g\, E \label{Mtree} \end{equation} 
\end{center}

With the same simple arguments, one can see that the vertex with four $\pi$s is proportional to $\epsilon_{step}g^2$, with five $\pi$s
to $\epsilon_{step}g^3$ and so on. Then we can list all the possible diagrams with three free legs and only one loop:

\vspace{2em}
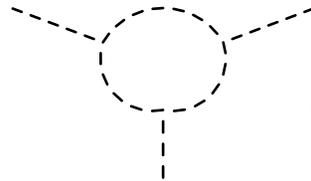
\begin{minipage}{.2\textwidth}
 \begin{center}
 \begin{fmffile}{3p1loop1}
 \begin{fmfchar*}(40,25)
 \fmftopn{pi}{2}
 \fmfbottom{pi3}
 \fmf{dashes}{pi1,v1}
 \fmf{dashes}{pi2,v2}
 \fmf{dashes}{pi3,v3}
 \fmf{dashes,left=0.57,tension=1/2}{v1,v2,v3,v1}
 \end{fmfchar*}
 \end{fmffile}
 \end{center}
\end{minipage}
\begin{minipage}{.255\textwidth}
\begin{center}
 \begin{equation} \mathcal{M}^{(1)} \sim 4\pi\,{\left(\epsilon_{step} \,g\, E\right)}^3 \end{equation} 
\end{center}
\end{minipage}

\vspace{2em}
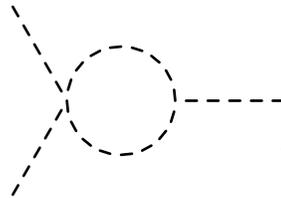
\begin{minipage}{.2\textwidth}
 \begin{center}
 \begin{fmffile}{3p1loop2}
 \begin{fmfchar*}(40,25)
 \fmfleftn{pi}{2}
 \fmfright{pi3}
 \fmf{dashes}{pi1,v1}
 \fmf{dashes}{pi2,v1}
 \fmf{dashes}{pi3,v2}
 \fmf{dashes,right,tension=1/2}{v1,v2,v1}
 \end{fmfchar*}
 \end{fmffile}
 \end{center}
\end{minipage}
\begin{minipage}{.255\textwidth}
\vspace{2em}
\begin{center}
 \begin{equation} \mathcal{M}^{(1)} \sim 4\pi\,\epsilon_{step}^2 \,{\left(g\, E\right)}^3 \end{equation}                                                                                         
\end{center}
\end{minipage}

\vspace{2em}
\begin{minipage}{.2\textwidth}
 \begin{center}
 \begin{fmffile}{3p1loop3}
 \begin{fmfchar*}(40,25)
 \fmftop{pi1}
 \fmfleft{pi2}
 \fmfbottom{pi3}
 \fmf{dashes}{pi1,v}
 \fmf{dashes}{pi2,v}
 \fmf{dashes}{pi3,v}
 \fmf{dashes,tension=1/2}{v,v}
 \end{fmfchar*}
 \end{fmffile}
 \end{center}
\end{minipage}
\begin{minipage}{.255\textwidth}
\begin{center}
 \begin{equation} \mathcal{M}^{(1)} \sim 4\pi\,\epsilon_{step} \,{\left(g\, E\right)}^3 \label {Mloop}\end{equation} 
\end{center}
\end{minipage}

The list ends here, as there are no more ways to connect three free legs with only one loop. Notice also that the largest effect comes
from the last diagram, where one has the lower power of $\epsilon_{step}$ and the higher power of $\beta$
(as $\epsilon_{step}\lesssim1$ and $\beta\gg1$).
Now, we can compare the tree-level amplitude with the loop contributions: the energy scale where the first one is comparable to the second, i.e.
\begin{equation}
 \mathcal{M}^{(0)}\sim\mathcal{M}^{(1)}\;,
\end{equation}
is to be considered as the maximum energy, $\Lambda$, at which the loop expansion is under control. Beyond that, interactions become
strongly coupled and the effective theory becomes non-unitary. It is easy to obtain $\Lambda$ from the previous equation, using
eqs. \eqref{Mtree} and \eqref{Mloop}:
\begin{equation}
 \Lambda^2\simeq16\pi{\left(\frac{M_{Pl}\sqrt{2\epsilon }}{\beta}\right)}^2 \; .
\end{equation}
If we want to trust our predictions, we should then make sure that the energy scales we study are all below this cut-off\footnote{The
same happens for resonant models, where one requires that the frequency of the resonance is smaller than the UV cut-off of the effective
theory \cite{Behbahani2011}.}. In particular,
\begin{equation} \label{beta_bound}
 \beta H\ll\Lambda \qquad \Longrightarrow \qquad \beta^2\ll\frac{2}{\sqrt{\pi}}\mathcal{P}_{\zeta,0}^{-1/2} \; .
\end{equation}

Some comments are in order. The bound \eqref{beta_bound} is very strict and should be taken with care, even from an observational point
of view. Indeed, from \emph{Planck} data analysis, the best fit of the power spectrum
seems to prefer very sharp features \cite{Planck_ps, Hu2013}, with $\beta \simeq 300$. However this is already out of the
allowed region, as from \eqref{beta_bound} we have $\beta\lesssim160$. This put serious questions on the consistency of these models
for those values of $\beta$, as we have shown that problems with the unitarity of the theory then arise.

Beyond the simplest case, with no other coefficients in the action but $\dot{H}(t)$, the EFTI naturally contains higher order operators,
which induce a speed of sound $c_s<1$ and are source of non-Gaussianity. These interactions will have a new UV cutoff
\cite{Creminelli2007},
\begin{equation}
 \Lambda_{c_s}^4\simeq16\pi^2M_{Pl}^2\dot{H}c_s^5 \; ,
\end{equation}
Then, it can be seen that there is an even stronger upper bound on $\beta$ requiring $\beta H$ be below this cutoff:
\begin{equation}\label{beta_bound_cs}
 \beta^2\lesssim c_s^2\,\mathcal{P}_{\zeta}^{-1/2} \; .
\end{equation}

This conclusion is very general and applies to every models where the slow-time dependence of the slow-roll parameters, the speed of sound
or any coefficient in the effective action is broken by some temporary effects with a characteristic time scale $\Delta t=b=1/\beta H$.
Physically, this bound is just telling us that we cannot ``effectively'' consider features on arbitrary small time scales, as the theory
of fluctuations is no more weakly coupled and perturbative unitarity is lost.

\section{Conclusions}

In this note, we considered the perturbative consistency of inflationary models with features by means of effective field theory methods.
By estimating the size of loop contributions to the $n$-point functions and comparing them with the tree-level computation, one can identify
the maximal energy scale at which the theory is unitary and perturbativity is safe. Then, by requiring that all the relevant energy scale
of the physics we are interested in are below this UV cutoff, we can derive bounds on the parameters of the models. While in the standard
slow-roll models of inflation, the only relevant energy scale is $H$, when features (or resonances) are present
there is a new energy scale $E\simeq1/\Delta t$ corresponding to the inverse of the characteristic time-scale of the interaction.
In the case of feature models, that we have studied here, we estimated the size of one-loop contributions to the three-point functions
and compared them to the tree-level expectation. Our main result is that \emph{there is a very strong upper bound on the sharpness
$\beta=1/\Delta tH$ of the feature, eq. \eqref{beta_bound_cs}, beyond which the unitarity of the theory is lost}.
The result is general and valid for any feature either in the slow-roll parameters, the speed of sound or any other coefficient in the
effective action for scalar perturbations during inflation. We point out that this bound is relevant even on observational grounds,
as current best-fits of the CMB power spectrum seem to prefer sharp feature models with a sharpness $\beta\simeq300$, which is already
outside our estimate of the unitarity bound.

\begin{acknowledgments}
D.C. thank Massimo Pietroni, Denise Vicino and Vicente Atal for useful discussions during the first stages of this work.
The work of N.B. and S.M. was partially supported by the ASI/INAF Agreement I/072/09/0 for the Planck LFI Activity of Phase E2.
\end{acknowledgments}

\vspace{1.5em}
{\bfseries Note Added: } While writing up this work, the paper arXiv:1402.1677 appeared, where similar issues are addressed.

\clearpage

\bibliography{bibliography}

\end{document}